\documentclass[a4paper]{toledoStyle}
\usepackage{epsfig}

\def\cc{{\rm cm ^{-3}}}
\def\lo{{\rm L_{\rm [OI]}}}
\def\llcii{{\rm L_{\rm [CII]}}}
\def\lfir{{\rm L_{\rm FIR}}}
\def\mum{\mu {\rm m}}
\def\rcof{(\llcii+\lo)/\lfir}
\def\6f{{\rm F_\nu(60 \mum)}}
\def\f100{{\rm F_\nu(100 \mum)}}

\def\i6{{\rm L(60 \mum)}}
\def\l1{{\rm L(100 \mum)}}
\def\r61{{\6f/\f100}}
\def\lo{{\rm L_{\rm [OI]}}}
\def\rcof{(\llcii+\lo)/\lfir}

\def\7f{{\rm F_{7 \mu m}}}

\def\o3f{\rm L_{\rm [OIII]}/\lfir}
\def\rat{{\rm L_{\rm [CII]}}/{\rm L_{\rm FIR}}}
\def\rcof{(\llcii+\lo)/\lfir}

\begin{document}

\title{Far Infrared Spectroscopy of star-forming galaxies: Expectations for the Herschel Space Observatory}

\author{Sangeeta Malhotra}

\institute{The Johns Hopkins University}

\maketitle 

\begin{abstract}

ISO has enabled far-infrared spectroscopy of a variety of galaxies.
Using the [CII] ($158 \mum$) and [OI]($63 \mum$) lines, we can
characterize the physical conditions in the star-forming ISM. These
observations also form the basis of our expectations for what the
Herschel Space Observatory will observe for high redshift
galaxies. While [CII] is suppressed in ULIRGs and normal galaxies with
high dust temperatures, it is stronger than expected in  metal poor
galaxies by factors of a few. Young galaxies at high redshifts might be
expected to be both metal poor and actively star-forming, leading to
contrary expectations for the [CII] line strength. The best prediction
for [CII] detection is derived by using the observed proportionality
between [CII] and mid-IR emission from PAHs. Using the observed [CII]/7
$\mum$ ratio and number counts from ISO deep surveys we predict that
HSO will be able to detect 100 sources/square-degree in the [CII] line.

\end{abstract}

\section{Introduction}

Herschel Space Observatory (HSO) will have the wavelength coverage and
sensitivity to detect high redshift galaxies at the peak of their dust
continuum emission.  The three instruments onboard HSO will also carry
out spectroscopy in the far-infrared and sub-millimeter. From the
FIR lines of $C^+$ and O, we can derive physical conditions in
the star-forming ISM at these redshifts. From optical observations it
seems that star-formation peaks at z=1-2 (e.g. Madau et al. 1996). It
would be interesting to investigate the properties of the ISM in
star-forming galaxies at these redshifts, and thus learn the causes and
consequences of higher star-formation.

In this paper I use ISO observations of the fine structure lines in
nearby galaxies to (a) summarize what we have learned about the ISM of
galaxies from ISO and (b) do feasibility calculations about what HSO
will see at high redshifts.

\section{The [CII] ($158 \mum$) line}

[CII] ($158 \mum$) line is the dominant coolant of the neutral ISM in
all but the hottest galaxies. For most of the observed galaxies
0.1-1\% of the FIR continuum emerges in this single line. Not
coincidentally, it is the best studied line in this wavelength regime.
It has been observed for normal galaxies (Stacey et al. 1991, Malhotra
et al. 1997, 1999, 2001, Leech et al. 1999, Pierini et al. 1999),
irregular/dwarfs (Smith et al. 1997, Bergvall et al. 2000, Hunter et
al. 2001, Madden et al. 2001), Ellipticals (Malhotra et al. 2000,
Unger et al. 2000) and luminous and ultraluminous infra-red galaxies
(Luhman et al. 1998, Fischer et al. 2001). So now we have a low
redshift ``basis set'' of various types of galaxies to draw upon.

The deficiency in [CII]/FIR in ULIRGs and normal galaxies (Malhotra et
al. 1997, Luhman et al. 1998) came as a surprise to many. The decrease
in [CII]/FIR correlates best with the IRAS colors of galaxies $\r61$
(Figure 1; Malhotra et al. 2001). It also correlates, but less
strongly, with other quantities like FIR/Blue colors and Infrared
luminosity. Since $\r61$, FIR/Blue and FIR luminosity correlate with
each other, we cannot say which correlation with [CII]/FIR is primary
and which ones are secondary. Luminous and ultraluminous galaxies from
the sample of Luhman et al. 1998 follow the same trends.  Two high
redshift quasars BRI~1202-0752 (z=4.69) and BRI~1335-0415 (z=4.41)
have measured upper limits on the [CII] flux and also follow the same
trendd (Benford 1999).

\begin{figure}[ht]
  \begin{center}
    \epsfig{file=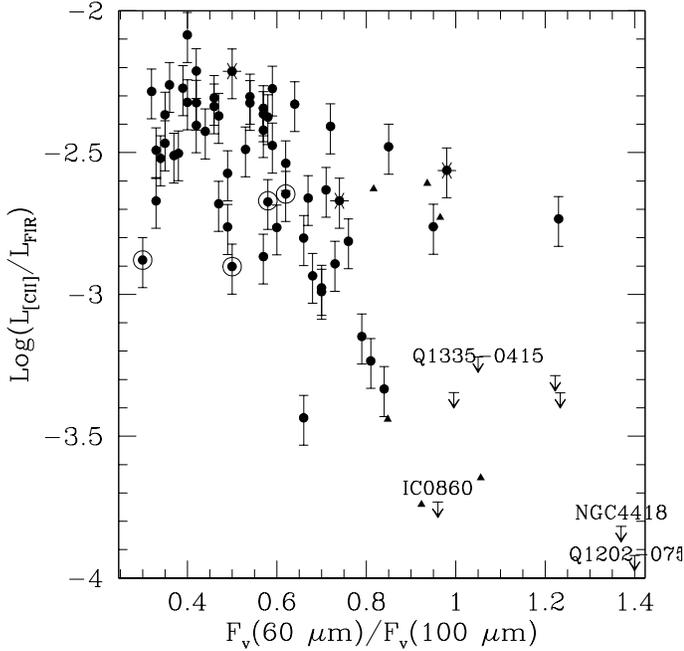, width=0.5\textwidth}
  \end{center}
\caption{The ratio of [CII] to far-infrared continuum, $\rat$, is
plotted against the ratio of flux in the IRAS $60 \mum$ and $100 \mum$
bands, $\r61$.  Filled circles are normal galaxies from the ISO-Key
Project sample.  Irregular galaxies are denoted with the star sign and
ellipticals with bulls eye symbols. Triangles are luminous and
ultraluminous galaxies from the sample of Luhman et al. (1998). There
is a trend for galaxies with higher $\r61$ (indicating warmer dust) to
have lower $\rat$, for normal as well as ULIRGs. Two normal galaxies
in a sample of 60 have no detected [CII], and they are identified with
labels and shown as upper limit symbols; other upper limits come from
Luhman et al. (1998). Rank correlation tests show that $\rat$ and
$\r61$ are correlated at the 4.4$\sigma$ level. $\rat$ upper limits
for two high redshift sources also follow this trend (Benford 1999).}
\end{figure}

\section{Irregular galaxies}

While FIR colors $\r61$ show the strongest correlation with $\rat$,
there seems to be a second parameter, which is apparent from Figure
1. Irregular galaxies have a higher $\rat$ and Ellipticals have lower
$\rat$ ratio.  It is not yet clear whether this has to do with the
lower metallicity affecting the chemistry of the ISM directly or
because low metallicity stellar populations produce a harder radiation
field. Hunter et al. 2001 and Madden et al. 2001 discuss in detail how
both $\rat$ and [CII]/MIR emission is higher for irregular galaxies.

Even more dramatic is the fact that [OIII] ($88 \mum$) line from HII
regions is very bright in 2 of our irregular galaxies. In IC~4662 and
NGC~1569 [OIII]/FIR is fully 1\% (Malhotra et al. 2001). This is
promising for detection of [OIII] at high redshifts, but it also means
that we cannot reliably assume that that the brightest (and sometimes
the only) line detected is [CII]. 

NGC~1569 and IC~4662 are dwarf galaxies and hence not very luminous. We
would not be able to see thier counterparts at z=1 or even z=0.5. The
[OIII] (88$\mum$) line from the metal poor irregular Haro 11 (Bergvall
et al. 2000) would be observable at z=1, even as the [CII] line falls
short. 

\section{FIR spectroscopy at high redshifts}

In figure 2 we compare the luminosity distribution of three prominent
FIR lines in the ISO key-project sample. These lines are [CII] (158
$\mum$), [OI] (63 $\mum$) and [OIII] (88 $\mum$). This is not a
luminosity function, since our sample is not volume limited, but
serves to compare the luminosity distribution among the three
brightest lines. We see that in spite of the [CII] deficiency which
cuts off the high luminosity tail of the distribution, [CII] is
clearly the more luminous of these three lines, and therefore offers
the best prospects for being widely detected.

\begin{figure}[ht]
  \begin{center}
    \epsfig{file=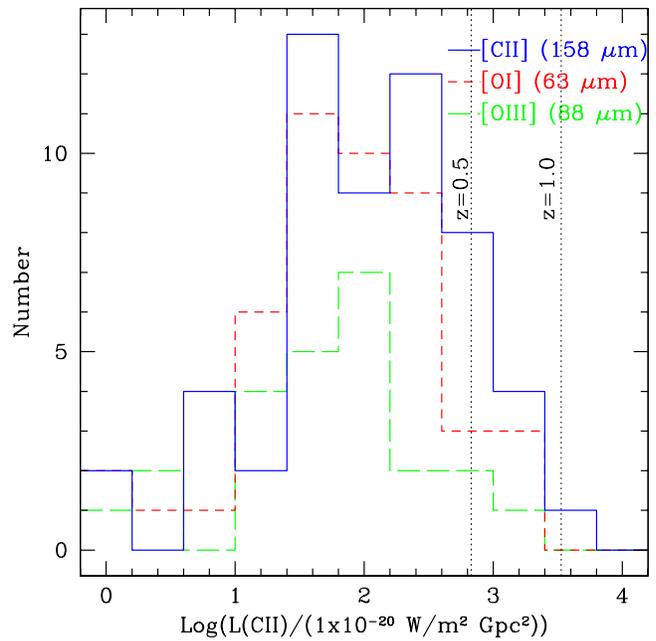, width=0.5\textwidth}
  \end{center}
\caption{ The luminosity distribution of the three prominent FIR lines
[CII] (158 $\mum$), [OI] (63 $\mum$) and [OIII] (88 $\mum$). The
sample here is the ISO Key-Project on normal galaxies. Since this
sample was designed to span the possible galaxy properties rather than
be a representative or a volume limited sample, this figure does not
show the luminosity function in any of the lines. However it is
instructive to see the relative luminosity distribution in the lines
and to note that [CII] is still the most luminous and likely to be
detected at redshifts 0.5-1. The x-axis shows the log of the fluxes
(in units of $1\times 10^{-20} W/m^2$) of these galaxies if they were
at a distance of 1 Gpc }
\end{figure}

The [CII] deficiency in the most active and luminous galaxies flattens
the luminosity function of the line relative to continuum. Since the
most luminous objects (e.g. ULIRGs) are deficient in [CII] the
prospects for detecting them at redshifts $\approx 4$ and higher are
dim. Observations of known high redshift objects with JCMT and CSO
have only yielded upper limits (van der Werf et al. 1998, Benford
1999). These upper limits are consistent with the trend seen between
$\rat$ and $\r61$ (Figure 1). It seems that the correlation between
$\rat$ and $\r61$ is stronger than between $\rat$ and luminosity
(Malhotra et al. 2001). In principle, a cool but luminous source would
be detectable at high redshifts. In practice, however, luminous
sources also tend to have warm FIR colors.  The [CII] deficiency in
high redshift galaxies may be somewhat mitigated if starburst galaxies
are metal poor, since metal poor galaxies have [CII]/FIR higher by a
factor of a few. But this effect is not expected to be large.

\subsection{Source density of high-z [CII] targets}

While the [CII]/FIR behaviour is complicated and correlates best with
$\r61$, we can exploit the [CII]/MIR constancy to predict the number
of galaxies that are observable with Herschel Space Observatory.

Helou et al. 2001 point out that the ratio of [CII] flux and the flux
in the mid-infrared band at 7 $\mum$ remains constant with FIR colors
and shows a smaller scatter than does $\rat$. The physical
interpretation of the correlation between [CII] and 7$\mum$ flux is
simple. The mid-infrared flux in the 6.75 band of ISO-CAM is dominated
by emission from Polycyclic Aromatic Hydrocarbons (PAHs).  We interpret the
stable [CII]/$\7f$ ratio as evidence that gas heating is dominated by
the PAHs or small grains which are also AFE carriers. 

Another interpretation of the constancy of [CII]/$\7f$ is that
 PAHs/very small grains and $C^+$ are co-extensive in PDR regions. The
 decrease in both [CII]/FIR (Malhotra et al. 1997, 2001) and $\7f$/FIR
 with F60/F100 (Helou et al. 2001, Lu et al. 2001) are then due to a
 smaller fraction of FIR arising from the PDR phase.

Regardless of the right explanation/interpretation of the
proportionality between [CII] and PAH emission, we can use it to
predict what the Herschel Space Observatory will be able to observe.

From the ISO key-project sample we derive 

$$[CII]=(\7f/1 mJy) \times 10^{-17.2} W/m^2$$

A sensitivity of $6 \times 10^{-18} W/m^2$ (HIFI) in the [CII]
line then translates to rest-frame 7 micron flux of 1 mJy.  Deep
observations with ISOCAM (Elbaz et al. 1999) show that the surface
density of sources with 15 $\mum$ fluxes $ \ge 1 mJy $ is roughly 100
per square-degree.

\begin{figure}[ht]
  \begin{center}
    \epsfig{file=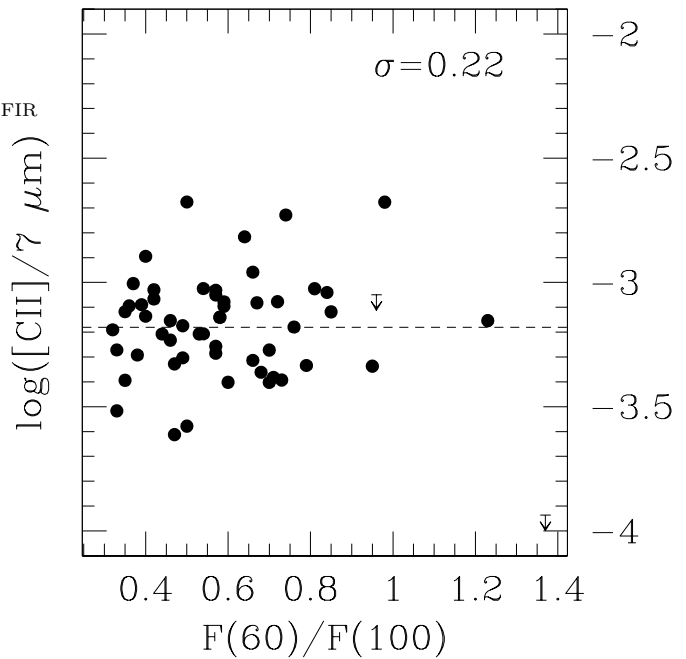, width=0.5\textwidth}
  \end{center}
\caption{ The ratio of [CII] flux and 7 $\mum$ flux is unvarying with FIR colors and shows a smaller scatter than $\rat$ ($\sigma ([CII]/\7f)= 0.22$. The [CII] flux is in units of $10^{-14} W/m^2$ and the 7 $\mum$ flux is in mJy.}
\end{figure}

There are two caveats we need to keep in mind when using the deep
counts in the mid-IR to predict the number of galaxies which will have
detectable [CII] flux. The first is that the flux in the mid-infrared
is dominated by spectral features due to aromatic molecules, which
means that as we get to higher redshifts K-corrections for fluxes
measured in any broad-band filter are substantial and not always of
the same sign (Xu et al. 1998). The other concern has to do with metal
poor galaxies. They show a deficiency in PAH features in the mid-IR
(Thuan et al. 1999), but also show higher [CII]/MIR and [CII]/FIR by
factors of 2-3. In these galaxies presumably most of the heating is
due to the very small grain component of dust, which emit in the
mid-IR but do not produce aromatic features.

\section{Physical conditions in the PDRs}

We derive the average physical conditions in the neutral gas in
galaxies, by comparing the observed line to continuum ratio $\rcof$
and $\r61$ with PDR models (e.g. Kaufman et al. 1999). The main
results of this study are (Malhotra et al. 2001):

(1) The derived temperatures at the PDR surfaces range from
270-900 K, and the pressures range from $ 6 \times 10^4 - 1.5 \times
10^7 {\rm K} \cc$. The lower value of the pressure range is
roughly twice the local solar neighborhood value and the upper end is
comparable to pressures in HII regions in starburst galaxies (Heckman,
Armus \& Miley 1990) which also corresponds to the pressure and
surface brightness at which starbursts saturate (Meurer et al. 1997).

(2) The average FUV flux $G_0$ and gas density $n$ scale as G$_0 \propto
n^{1.4}$. This correlation is most naturally explained as arising from
Str\"{o}mgren sphere scalings if much of the line and continuum
luminosity arises near star-forming regions.  From simple
Str\"{o}mgren sphere calculations (cf. Spitzer 1978) we can derive
that the FUV flux at the neutral surface just outside the
Str\"{o}mgren sphere should scale as G$_0 \propto n^{4/3}$, which is
consistent (within errors) to the scaling seen in Figure 4. The $G_0$,
$n$ and $P$ which we derive for given galaxy represent a luminosity
weighted average value and mostly represents dense GMCs which lie
close to the OB stars. 

(3) The range of $G_0$, $n$ and $P$ which we derive for the different
galaxies can reflect several interesting differences in their star
formation processes and histories.  If global star formation is
episodic then high G$_0$ and $n$ imply that the galaxy is observed
shortly after a burst because the OB stars have not moved far from
their natal clouds.  Alternatively, the differences in $G_0$, $n$ and
$P$ may reflect differences from galaxy to galaxy in the properties of
the GMCs which form the OB stars. Larger GMCs may keep their OB stars
embedded for a longer fraction of their lifetime, resulting in higher
average $G_0$ and $n$. Or the GMCs could be the same size but denser,
so that the higher density ambient gas would lead to smaller Stromgren
spheres and higher $G_0$ at the edges of the spheres.

\begin{figure}[ht]
  \begin{center}
    \epsfig{file=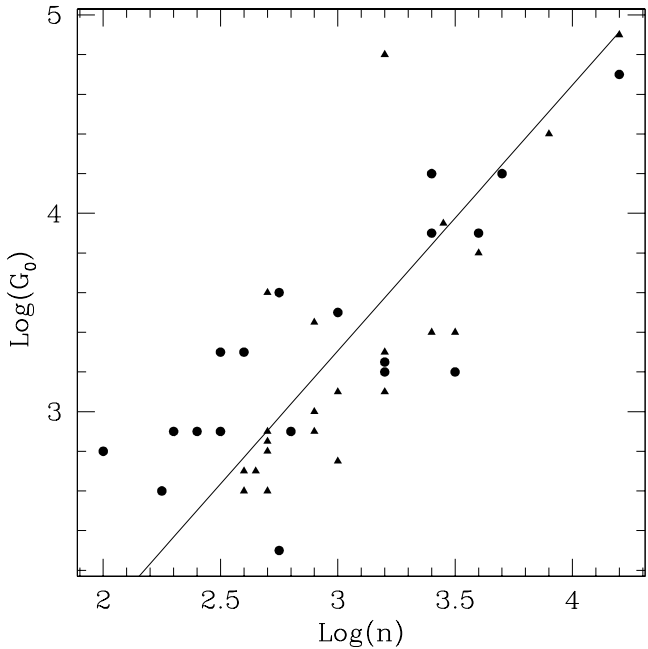, width=0.5\textwidth}
  \end{center}
\caption{This figure shows the derived far-UV flux G$_0$ and gas
density $n$ solution for the ISO key-project galaxies based on
comparison of FIR line data and PDR models of Kaufman et al. 2001. A
least square fit is made to the $G_0$ vs. $n$ relation assuming equal
error in both axes.  The best fit slope is 1.4, i.e. G$_0 \propto
n^{1.4}$, which is consistent with PDRs surrounding ionization bounded
expanding HII regions.}
\end{figure}

\section{Conclusions}

It seems unlikely that spectroscopy with HSO will be used to find
redshifts of infrared bright galaxies with unknown or obscured optical
counterparts simply because scanning the possible redshift range would
be time consuming. But once the redshifts are known, one can use the
observations of [CII] and [OI](63 $\mum$) lines to derive the physical
conditions in PDRs. Or if photometric redshifts using mid to
far-infrared colors can be refined enough, one could search for fine
structure lines to get the precise redshifts.

FIR spectroscopy of modest redshift ( $z \simeq 1$) galaxies would be
feasible with the HSO.  This would be valuable to understand the
higher global rates of star-formation in this epoch at roughly half
the age of the universe. With ISO we have been able to characterize the 
ISM in star-forming local galaxies, so we do have a comparison set. 

\acknowledgements

I would like to thank my collaborators on the ISO Key Project - George
Helou, Michael Kaufman, David Hollenbach, Danny Dale, Alexandra
Contursi and Gordon Stacey. SM's research funding is provided by NASA
through Hubble Fellowship grant \# HF-01111.01-98A from the Space
Telescope Science Institute, which is operated by the Association of
Universities for Research in Astronomy, Inc., under NASA contract
NAS5-26555.

\end{document}